\documentclass[preprint,12pt]{elsarticle}

\usepackage{amssymb}
\usepackage{amsmath}
\usepackage{graphicx}
\usepackage{color}
\usepackage{longtable}
\biboptions{sort&compress}

\journal{Journal of Solid State chemistry}

\begin{document}

\begin{frontmatter}

%% Title, authors and addresses

%% use the tnoteref command within \title for footnotes;
%% use the tnotetext command for the associated footnote;
%% use the fnref command within \author or \address for footnotes;
%% use the fntext command for the associated footnote;
%% use the corref command within \author for corresponding author footnotes;
%% use the cortext command for the associated footnote;
%% use the ead command for the email address,
%% and the form \ead[url] for the home page:
%%
%% \title{Title\tnoteref{label1}}
%% \tnotetext[label1]{}
%% \author{Name\corref{cor1}\fnref{label2}}
%% \ead{email address}
%% \ead[url]{home page}
%% \fntext[label2]{}
%% \cortext[cor1]{}
%% \address{Address\fnref{label3}}
%% \fntext[label3]{}

\title{Magnetic Behaviour of the Bi$_{2-y}$Sr$_{y}$Ir$_{2}$O$_{7}$ Pyrochlore Solid Solution}

%% use optional labels to link authors explicitly to addresses:
%% \author[label1,label2]{<author name>}
%% \address[label1]{<address>}
%% \address[label2]{<address>}

\author[a]{C. Cosio-Castaneda}
\author[b]{P. de la Mora}
\author[c]{F. Morales}
\author[c]{R. Escudero}
\author[a]{G. Tavizon\corref{cor1}}
\ead{gtavizon@unam.mx}

\address[a]{Departamento de F\'{i}sica y Qu\'{i}mica Te\'{o}rica, Facultad de Qu\'{i}mica}
\address[b]{ Departamento de F\'{i}sica, Facultad de Ciencias} \address[c]{Instituto de Investigaciones en Materiales\\

Universidad Nacional Aut\'{o}noma de M\'{e}xico, Ciudad Universitaria, M\'{e}xico D. F. 04510, M\'{e}xico}

\cortext[cor1]{Corresponding author}

\begin{abstract}
The temperature dependence of magnetic susceptibility of the Bi$_{2-y}$Sr$_{y}$Ir$_{2}$O$_{7}$ system was studied from 2 to 300 K. According to these measurements, the solid solution $(0\leq y\leq 0.9)$ does not show any magnetic transition; however, a noticeable deviation from the Curie-Weiss law occurs and the magnetic behavior of this system can be ascribed to short-range magnetic order that is also present in several geometrically frustrated systems. The Pauli magnetic susceptibility decreases as the Sr content increases, this can be associated with a shift towards Ir$^{5+}$ in samples; also the Ir effective magnetic moment decreases. The estimated Wilson-ratio indicates the presence of strong electron-electron correlations.
\end{abstract}

\begin{keyword}
Pyrochlore oxides\sep Iridiates\sep Short-range magnetic order.

%% MSC codes here, in the form: \MSC code \sep code
%% or \MSC[2008] code \sep code (2000 is the default)

\end{keyword}

\end{frontmatter}

%%
%% Start line numbering here if you want
%%
% \linenumbers

%% main text

\section{Introduction}

The Iridium  $\alpha$-pyrochlore family of compounds, A$_2$Ir$_2$O$_7$, spans a broad spectrum of electrical and magnetic properties that make them prime candidates for the exploration of complex phenomena in solid state chemistry and physics \cite{1}. The crystalline structure of this family is face-centered cubic with space group $Fd\bar{3}m$ (S. G. 227), that can be described as two interpenetrated sublattices, A$_2$O2 and Ir$_2$O1$_6$. In A$_2$Ir$_2$O$_7$ the Ir atoms form the 3D equivalent of a Kagome structure. When the A cation is a rare-earth (RE) element, the effective magnetic moment agrees with the expected value for the RE$^{3+}$ ions, indicating its localized nature \cite{2}. All of the A$_2$Ir$_2$O$_7$ (A=RE) compounds have Weiss temperatures $\theta$ corresponding to an antiferromagnetic (AF) order and none of their moments exhibits magnetic ordering at temperatures well below the $\theta$ value (down to 0.3 K) \cite{2}. Systems with a nonmagnetic A-cation have been reported; Y$_2$Ir$_2$O$_7$ is a Mott insulator with a small ferromagnetic component below 170 K, as a consequence of either a spin-glass ordering or spin canting \cite{4}. When Y is partially substituted by Bi, Y$_{2-x}$Bi$_{x}$Ir$_2$O$_7$, experimental data of transport properties indicate a metal-insulator transition at $x\sim$ 0.5 \cite{5}. In the solid solution Y$_{2-x}$Ca$_x$Ir$_2$O$_7$, in which Y and Ca are nonmagnetic but have different oxidation states, partial substitution of Y changes the electrical and magnetic properties substantially; a metallic behavior appears when $x\geq 0.3$, also the ferromagnetism (transition temperature at 170 K) disappears \cite{4}.

Recently weak magnetic frustration has been proposed to occur in Eu$_2$Ir$_2$O$_7$ \cite{6}. This compound presents an interesting case of a magnetic frustrated system, in which  Eu$^{3+}$ is a $J=0\ (L=S)$ nonmagnetic cation and only the Ir$^{4+}\ 5d$ electrons contribute to magnetism in this compound, in which a metal-insulator/AF transition is reported at 120 K \cite{3}. The occurrence of localized magnetic moments and a $5d$ cation is an unexpected situation, since the Ir$^{4+}$ wave functions should result in a metallic conduction via Ir-derived bands \cite{6}. Poor metals such as RInCu$_4$ (R=Gd, Dy, Ho) \cite{7}  and the Cd doped system (RIn$_{1-x}$Cd$_x$Cu$_4$, R=Gd-Tm) \cite{8}, behave as frustrated systems. In the latter system, long-range interactions  introduced into the system by the conduction electrons are thought that they reduce the magnetic frustration. In itinerant electron systems the exchange interactions are not restricted to nearest neighbors and magnetic order can occur if the magnetic interactions are large enough; in this situation the mechanism is via the RKKY interaction \cite{9}.

The Bi$_2$Ir$_2$O$_7\ \alpha$-pyrochlore was first reported by Bouchard and Gillon \cite{10} and later structurally analyzed by B. J. Kennedy \cite{11}. This is a metallic system in which no magnetic transition has been reported. The temperature dependence of the magnetic susceptibility is Curie-Weiss-like (CW), however a detailed analysis that explains the magnetic behavior has not been completely done. The metallic character in the Y$_{2-y}$Bi$_y$Ir$_2$O$_7$ system shows a Bi-content dependence \cite{5,12}, and we recently showed \cite{13} that in this type of compounds the main contributors at the Fermi energy $(E_F)$ are Ir and O1 atoms. In the present work we report the effect of Bi substitution by Sr on the magnetic behavior of Bi$_2$Ir$_2$O$_7$ (Bi$_{2-y}$Sr$_y$Ir$_2$O$_7,\ 0\leq y\leq 0.9)$. This substitution is also accompanied by important changes in the local geometry around Ir \cite{13}. The system studied here shows an enhanced paramagnetism that can be described by a modified Curie-Weiss (mCW) equation in which short-range magnetic interactions are taken into account. On the other hand, the Pauli magnetic susceptibility decreases with the Sr content and this should be indicative of a similar tendency of the electronic density of states at the Fermi energy $N(E_F)$.

\section{Experimental details}

Polycrystalline samples of Bi$_{2-y}$Sr$_y$Ir$_2$O$_7$ were prepared by solid state reaction with stoichiometric amounts of IrO$_2$, Bi$_2$O$_3$ and SrCO$_3$ powders. Details of syntheses and structure analysis can be found elsewhere \cite{13}. DC magnetization data were acquired with a SQUID magnetometer (MPMS, Quantum Design) in a temperature range between 2 and 300 K. Field Cooling mode measurements were performed under a magnetic field of 0.1 T for all samples.

\section{Results and discussion}
 In Bi$_{2-y}$Sr$_y$Ir$_2$O$_7$ a monotonous increase of the cubic lattice parameter is observed as a function of the strontium content, $y$, which is in agreement with the difference between the Sr$^{2+}$ and Bi$^{3+}$ ionic radii (Sr$^{2+} = 1.26$ {\AA} and Bi$^{3+} = 1.17$ \AA, both in 8-coordination \cite{15}. Additionally, a linear reduction of the Ir-O1 bond distance was observed as consequence of the steric effect associated with the increase in the oxidation state of the Ir cation and the repulsive effect due to the larger Sr radius \cite{13}. These structural modifications were earlier predicted by Koo et al. \cite{16}  in terms of the coordination environment of the O1 atom.

When a divalent cation is included into the A$^{3+}_{2}$B$^{4+}_{\ 2}$O$_7$  $\alpha$-pyrochlore lattice to form an aliovalent solid solution, a charge-compensation process is expected to occur by the appearance of a positive charged lattice defect \cite{17,18}. This charge-compensation can be considered through the Kr\"{o}ger-Vink notation \cite{19}. According to the chemical reaction, in the localized defects scheme, there are three possible consequences resulting from the Bi by Sr substitution:

\begin{equation}
\rm SrO + O^{x}_0 + Ir^{x}_{Ir} \xrightarrow[ ]{Bi_2Ir_2O_7} Sr''_{Ir} + V^{\bullet\bullet}_0 + IrO_2
\end{equation}

\begin{equation}
\rm 2SrO + O^{x}_0 + 2Bi^{x}_{Bi} \xrightarrow[ ]{Bi_2Ir_2O_7} 2Sr'_{Bi} + V^{\bullet\bullet}_0 + Bi_2O_3
\end{equation}

\begin{equation}
\rm SrO + Ir^{x}_{Ir} \xrightarrow[ ]{Bi_2Ir_2O_7} Sr'_{Bi} + O^{x}_0 + Ir^{\bullet}_{Ir}
\end{equation}

When Sr$^{2+}$ is incorporated into the Bi$_2$Ir$_2$O$_7$ lattice it could occupy an Ir$^{4+}$ or Bi$^{3+}$ site, with the consequent generation of an oxygen vacancy and the corresponding metal oxide (IrO$_2$ and Bi$_2$O$_3$, Schottky reactions 1 and 2). An additional effect related with the incorporation of Sr$^{2+}$ at the Bi$^{3+}$ site is associated with a change of the oxidation state of Ir (reaction 3). Considering the oxygen occupation values obtained from the Rietveld analysis \cite{13}, it is possible to think that Schottky reactions could happen, however the corresponding bismuth or iridium oxides were not found in the diffraction patterns as products of such reaction. It should be pointed out that the prolonged time of the solid-state reaction could lead to evaporation and disproportionation of the iridium-oxide, releasing oxygen \cite{20}. On the basis of the experimental results, it is clear that reaction 3 is being preferred but is difficult to accurately determine the oxygen content of samples.

\begin{figure}[bth]
\begin{center}
\includegraphics[scale=0.4] {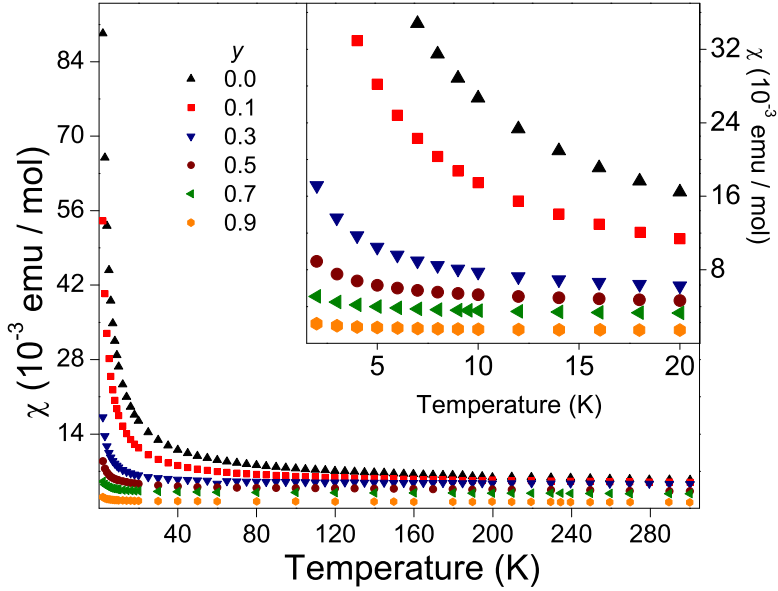}
\caption{\label{fig1}Magnetic susceptibility as a function of temperature of Bi$_{2-x}$Sr$_x$Ir$_2$O$_7$. The inset shows a magnification of the low temperature region.}
\end{center}
\end{figure}

The temperature dependence of the magnetic susceptibility of the Bi$_{2-y}$Sr$_y$Ir$_2$O$_7$ compounds with the $y$-compositions 0.0, 0.1, 0.3, 0.5, 0.7 and 0.9 are shown in figure \ref{fig1}. All susceptibility data were corrected for core diamagnetism \cite{21}. As can be observed in this figure, none of the samples showed a magnetic transition, at least down to 2 K, and the magnetic susceptibility diminishes as the Sr content increases. The magnetic susceptibility reduction can be explained in terms of a change in the oxidation state of the Ir atoms, since Ir is the only magnetic cation present in the Bi$_{2-y}$Sr$_y$Ir$_2$O$_7$ series.

The coordination environment around the Ir  cation in the $\alpha$-pyrochlore crystal lattice is a distorted octahedron in the form of a trigonal antiprism (TA) \cite{22}, that is, the lattice is distorted in the [111] direction. As a consequence, from the $t_{2g}$($d_{xy}, d_{yz}, d_{xz}$) and $e_g$  ($d_{x^{2}-y^{2}}, d_{z^{2}}$) regular octahedron splitting, now $t_{2g}$ splits into $e'_g (d_{x'z'}, d_{y'z'}$) and $a_{1g}$ $(d_{x'y'})$ \cite{23},  where $x', y'$ and $z'$ is a new coordinate system with $z'$ in the [111] direction. On the other hand, the bands are wide \cite{13} and the $e'_g-a_{1g}$ splitting for the studied system is considerably smaller. Thus, in the crystal field scheme, the electron configuration of Ir$^{4+}$ is [Xe]$4f^{14}5d^{5}:\ e_{g}^{'4}a_{1g}^{\ 1}$, with a spin magnetic moment $S = 1/2$. In this way the explanation of the magnetic susceptibility reduction as a function of the Sr  content is that the inclusion of Sr in  trivalent cation positions (Bi), produces a partial oxidation of iridium atoms from Ir$^{4+}$ to Ir$^{5+}$.

The Ir$^{5+}$ produced to achieve electroneutrality has an electron configuration [Xe]$4f^{14}5d^4: e_{g}^{' 4}a_{1g}^{\ 0}$, with $S=0$, therefore it would not contribute to the magnetic susceptibility \cite{24}, and the susceptibility of  the compounds is due to the remaining Ir$^{4+}$ \cite{25,26}; consequently, magnetic susceptibility of samples should diminish as is experimentally found. In contrast, in an octahedral crystal field splitting (low spin), Ir$^{4+}$ has $S = 1/2$ (the same that for TA splitting), but Ir$^{5+}$ has $S = 1$. The experimental measurement of magnetization seems to point to the existence of nonmagnetic Ir$^{5+}$, which increases as a function of the Sr content, see figure \ref{fig1}. Several other pyrochlore systems containing Ir$^{5+}$ have previously been reported, in which the Ir valence was stated by X-ray photoemission spectroscopy (XPS) \cite{27}  and X-ray absorption near edge spectroscopy (XANES) \cite{28}.

In order to explain in detail the magnetic behavior of the Bi$_{2-y}$Sr$_{y}$Ir$_2$O$_7$ system, a first approximation was done fitting the susceptibility data to the CW equation, $\chi (T)=C/(T-\theta)$. This fit was performed for the high temperature region but the $C$ and $\theta$ values are not congruent with those expected for a paramagnet.

All samples of the Bi$_{2-y}$Sr$_{y}$Ir$_2$O$_7$ series show metallic behavior, but at a fixed temperature, for a bigger Sr-content corresponds a larger resistivity. The metallic conductivity and non CW behavior is not exclusive of magnetic compounds crystallizing in the $\alpha$-pyrochlore structure, this has also been observed in iron pnictides \cite{31} and La$_{2-x}$Sr$_x$CuO$_4$ systems \cite{32}. Likewise, short-range magnetic interactions between rare earth ions on the A site of magnetic pyrochlores have been reported in R$_2$Ru$_2$O$_7$ (R=Tb, Er and Yb) \cite{33}.Therefore, it is necessary to consider short-range magnetic order and spin fluctuations \cite{29,30} in the interpretation of the magnetic behavior of Bi$_{2-y}$Sr$_{y}$Ir$_2$O$_7$.

\begin{figure}[bth]
\begin{center}
\includegraphics[scale=0.4]{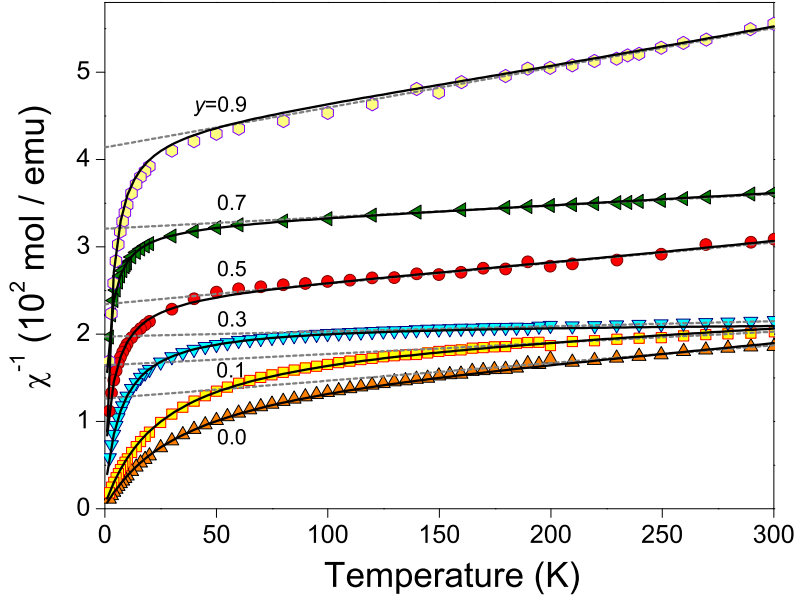}
\caption{\label{fig2}Inverse of the magnetic susceptibility $\chi^{-1}$ of the Bi$_{2-y}$Sr$_{y}$Ir$_2$O$_7$ as a function of temperature. The dashed lines represent a linear fitting of data above 100 K. The continuous lines represent the data fitting using the modified Curie-Weiss equation, see text.}
\end{center}
\end{figure}

Based on the theoretical expressions deduced by Opechowski and Li, in which the short-range magnetic order effects are included in power series of $Jz/2k_BT$ \cite{29,34,35}, a mCW equation was tried to fit the observed magnetic susceptibilities. The $Jz/2k_B$ term could be associated with the $\theta$  value in the Curie-Weiss model when $S = 1/2$, since  $\theta = 2zJS(S+1)/3k_B$, where $J$ is the exchange integral and $z$ is the spin coordination number. Additionally, a phenomenological linear temperature term $\alpha T$ was included. The modified CW equation is \cite{32,36};
\begin{equation}
\chi = \chi_0 +  \alpha T + \frac{C}{T\Big[1 + \big( \frac{\theta}{T}\big)^2 - \left( \frac{\theta}{T}\right)^3\Big]-\theta}
\label{mcw}
\end{equation}

The parameters obtained with the mCW fitting are shown in Table 1. It is worth mentioning here that when an equation such as $\chi = C/(T-\theta ) + \alpha T+\chi_0$ was tried, the fitting was not satisfactory, especially in the low temperature region, $T<$100 K. Even though all samples exhibit a paramagnetic behavior in the 2-300 K range, a noticeable  deviation from the linearity in  $\chi^{-1}$ is observed for the low temperature region. This departure goes to lower temperatures as the Sr content increases. The further from Ir$^{4+}$ towards Ir$^{5+}$, the larger is the fitting range to the CW equation.

\begin{table}
\caption{Magnetic parameters obtained from fitting of magnetic susceptibility data of the Bi$_{2-y}$Sr$_{y}$Ir$_2$O$_7$ system using equation \ref{mcw}.}

\begin{tabular}{ccccc}
\hline
$y$     &$\chi_0\ 10^{-3}$ & $\alpha\ 10^{-6}$ & $C$ &{$\theta$}\\
   & (emu/mol)& (emu/K-mol)& (emu/K-mol) & (K)\\ \hline
0.0 & 5.93 & - 4.85 & 0.2148 & - 0.45 \\
0.1 & 5.04 & - 2.08 & 0.1287 & - 0.49 \\
 0.2 & 4.70 & -0.09 & 0.0316 & - 0.43 \\
 0.3 & 4.41 & - 1.33 & 0.0151 & -0.60 \\
 0.4 & 4.37 & - 0.30 &0.0149 & - 0.60 \\
 0.5 & 4.00 & - 2.62 & 0.0134 & - 0.56 \\
 0.6 & 4.43 & - 0.37 & 0.0124 & - 0.39 \\
 0.7 & 3.06 & -1.04 & 0.0051 & - 0.43 \\
 0.8 & 1.98 & - 0.24 & 0.0040 & - 0.41 \\
 0.9 & 1.37 & - 0.67 & 0.0018 & - 0.35 \\
 \hline
\end{tabular}
\end{table}

The results of the magnetic susceptibility fitting obtained with the equation \ref{mcw} show that the $\alpha$  term is small when it is compared with $\chi_0$, but this phenomenological term is necessary to obtain a good adjustment of the experimental data and also permits a suitable interpretation of $C$ and $\theta$.

If the conventional procedure to determine $C$ and $\theta$  from magnetic susceptibility data had been used, that is, extrapolating  $\chi^{-1}$ $vs$. temperature data for the high temperature range, then one obtains very large $-\theta$  values: 470 K $(y=0)$; 1020 K $(y=0.1)$; 2940 K $(y=0.3)$; 1138 K $(y=0.5)$; 2687 K $(y=0.7)$ and 1577 K $(y=0.9)$. From the $C$ values, Ir effective magnetic moments of 5.7, 7.4, 11.18, 7.14, 10.11 and 5.67 Bohr magnetons $(\mu_B)$, are obtained, which are unacceptable (for the Ir$^{4+}$ free-ion $\mu_{eff}=1.73\ \mu_B$ \cite{52}).

\begin{figure}[tbh]
\begin{center}
\includegraphics[scale=0.4] {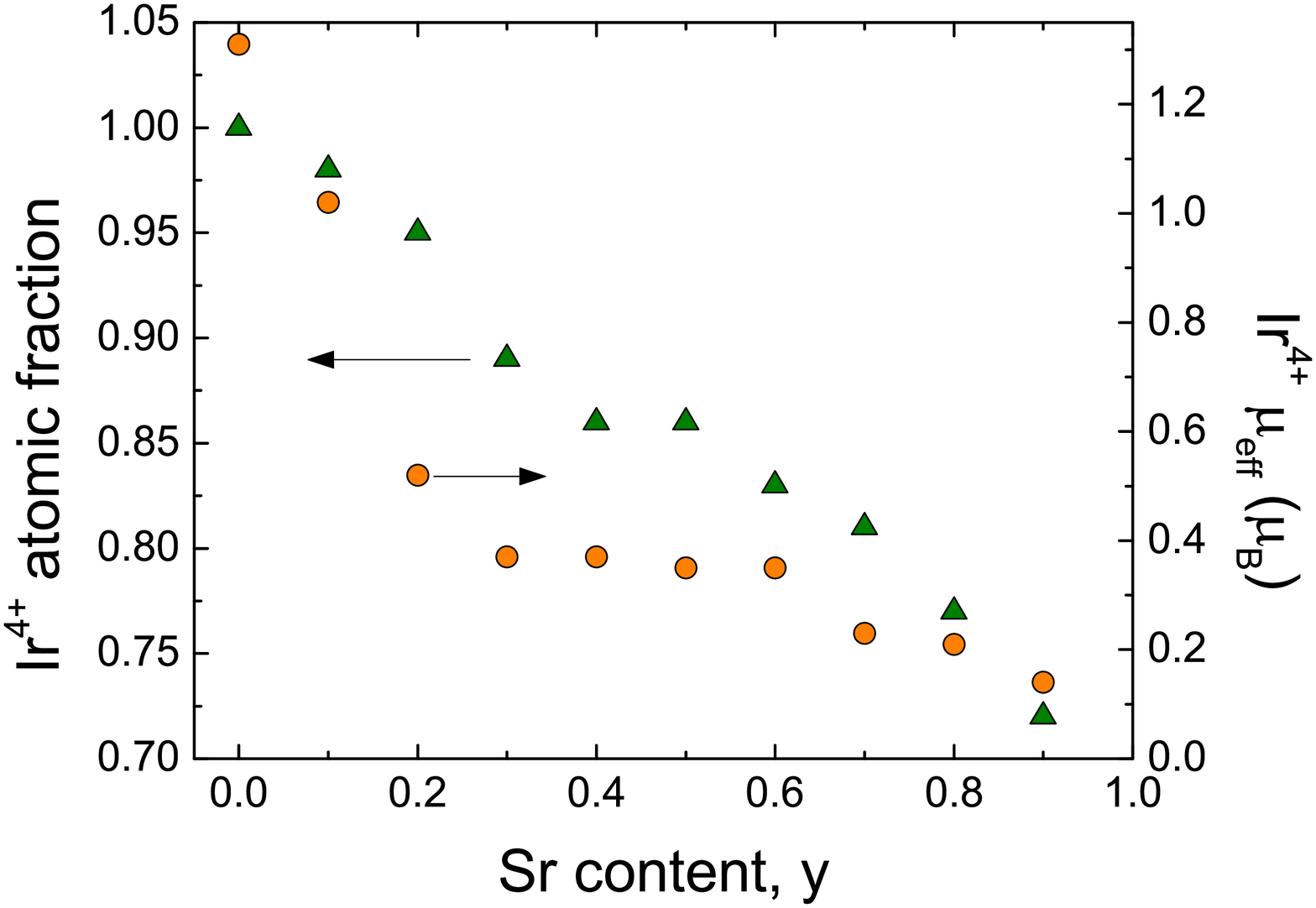}
\caption{\label{f3}Ir$^{4+}$ atomic fraction (triangles) and iridium effective magnetic moment, $\mu_{eff}$, (circles)  in the Bi$_{2-y}$Sr$_{y}$Ir$_2$O$_7$ solid solution as a function of the $Sr$ content.  The Ir$^{4+}$ atomic fraction is obtained from the $y$-dependence of the crystal cell-parameters, and the  $\mu_{eff}$ from the $C$ value, see table 1.}
\end{center}
\end{figure}

The small negative values obtained for $\theta$  with equation \ref{mcw} suggest a weak AF coupling in all samples of this solid solution. No magnetic transition  was observed, if this transition exists, it might occur at temperature below 2 K. Very recently, Qi et al. reported the magnetic and thermal behavior of Bi$_2$Ir$_2$O$_7$ single-crystal. They showed that in the range 0.05-350 K there are no anomalies indicating long-range order, and found that $\chi_0=1.07\times10^{-2}$ (emu/mole),  $\theta= -2.3$ K and a $\mu_{eff}=0.10\ \mu_B/$Ir and the magnetic susceptibility of that sample noticeably differs from that of the CW equation \cite{37}.

Since the cell-parameter of the Bi$_{2-y}$Sr$_{y}$Ir$_2$O$_7$ solid solution behaves according to the Vegard's law, then knowing the ionic radii of Bi$^{3+}$ (1.17{\AA},  CN=8 ), Sr$^{2+}$ (1.26{\AA}, CN=8), Ir$^{4+}$ (0.625{\AA}, CN=6) and Ir$^{5+}$ (0.57{\AA},  CN=6) (CN is the coordination number) \cite{15} an approximation of the Ir$^{5+}$ content can be obtained. This procedure is better than that using the Ir-O bond-distances \cite{13}, since these values are influenced by the effect of Bi substitution by Sr, mentioned above \cite{16}. Using this approximation the Ir$^{4+}$ atomic fractions, $x$, in samples were calculated.

The Ir$^{4+}$ atomic fractions reported in this work agree well with the oxygen occupation factors from previous Rietveld refinements results \cite{13}. The difference between the stoichiometric Sr value, $y$, and the Ir$^{5+}$ content of samples to preserve  electroneutrality, can be attributed to the oxygen deficiencies of samples \cite{20}.

 The $C$ term of the equation \ref{mcw} can be interpreted as the effective magnetic moment per magnetic ion, $\mu_{eff}$. According to equation \ref{mcw}, and taking into account the remaining Ir$^{4+}$ content, the $\mu_{eff}$ values are shown in figure \ref{f3}. For the $y = 0.0$  sample, $\mu_{eff}$ corresponds to 76\% of the expected value for $S = 1/2$ in a TA geometry, however, in Y$_2$Ir$_2$O$_7$, which is magnetically similar to Bi$_2$Ir$_2$O$_7$, Fukazawa et al. \cite{4} report 14\% of the expected value with $S = 1/2$. This difference could be associated with oxygen or iridium vacancies, because in both compounds the atomic distances Ir-O1 and bond angles Ir-O1-Ir are similar. Moreover, the $\mu_{eff}$ values of the Bi$_{2-y}$Sr$_{y}$Ir$_2$O$_7$ system indicate that with increasing the Sr$^{2+}$ content, the effective magnetic moment diminishes until it reaches 8.1\% of the expected value, at $y = 0.9$.

Ir-$\mu_{eff}$ values smaller than the expected value of 1.73 $\mu_B$ \cite{52} have been reported for several Ir compounds, these values range from 0.13 to 0.69 $\mu_B$ \cite{47}. One explanation of this small value of $\mu_{eff}$ is attributed to the strong spin-orbit coupling of $Ir$ cations and the strong exchange interactions that lower the magnetization of samples \cite{47}. In the case of Eu$_2$Ir$_2$O$_7$, muon spin rotation and relaxation experiments did not suffice to provide the magnetic structure of such compound, furthermore neutron scattering measurements result  prohibitively difficult because of the high neutron absorption cross section of Ir nucleus \cite{6}.

As can be observed from \ref{fig2}, the magnetic measurements the Bi$_{2-y}$Sr$_{y}$Ir$_2$O$_7$ solid solution, rather than being the result of a type of magnetic alignment of moments under an applied magnetic field (paramagnetic gas), they seem to show the effect of short-range magnetic interactions, which, prevailing at the whole 2-300 K temperature range, are not capable to condensate into a large-range AF state. In this way, while no noticeable variation of the $\theta$  value dependent of the Sr content is observed, there is a clear departure from the CW behavior that can be associated with short-range interactions, especially in the low temperature region.

For $y=$0 and 0.1 the calculated magnetic moments   are  1.23 and 1.02 $\mu_{B}$, respectively. These values agree well with those reported for Ir$^{4+}$ in Eu$_2$Ir$_2$O$_7$ (1.1 $\mu_{B}$) \cite{6}. On the other hand, the independent temperature susceptibility terms, $\chi_0$, obtained are of the order of $10^{-3}$ emu/mol, and these decrease as the Sr content increases. These $\chi_0$ values are similar to the values obtained for other pyrochlore iridates \cite{2,4} such as Y$_2$Ir$_2$O$_7$ and ruthenates, Pb$_2$Ru$_2$O$_{6.5}$, \cite{43,44}.

The tendency of $\chi_0$ to diminish with the Sr  content could have origin in two different processes. On one hand, one could think of a reduction of the $N(E_F)$ associated to the appearance of Ir$^{5+}$. However, according to previous results \cite{13} of DFT density of states calculation for Bi$_2$Ir$_2$O$_7$ and BiSrIr$_2$O$_7$, these  $N(E_F)$  values are practically insensitive to the Sr content. Since in both systems the main contributors to $N(E_F)$ are Ir and O1 (95\% in Bi$_2$Ir$_2$O$_7$ and 96.5\%  in BiSrIr$_2$O$_7$), Ir$^{5+}$ is not expected to be the main responsible for this diminishing in the  $\chi_0$  values. On the other hand, it is worth mentioning that, associated to the $5d$ electronic configuration and the local geometry of Ir$^{4+}$, a strong spin-orbit coupling (SOC) has been pointed out as responsible of spectacular electronic properties in localized \cite{6,45} and metallic systems \cite{2,46}. The existence of this strong SOC would significantly modify the $\chi_0$ values with respect to those observed in $3d$ and $4d$-transition metal oxides. At this regard the Wilson ratio, $R_W$, provides an empirical indicator of the strength of the electron-electron correlation.

As the case of the `5M' barium iridiate, BaIrO$_3$, the temperature independent magnetic susceptibility,  $\chi_0$, of Bi$_{2-y}$Sr$_{y}$Ir$_2$O$_7$ is in the range of the free electron system \cite{47}. In the free-electron approximation of metals the Pauli susceptibility, $\chi^P$ can be calculated from $\chi_0$ by subtracting the diamagnetic Landau contribution, $\chi^P = (3/2) \chi_0$  (where $\chi_0$ is corrected for core diamagnetism), and  $\chi^P$ is proportional to $N(E_F)$.

For Bi$_2$Ir$_2$O$_7$ single-crystal in the low temperature regime (50 mK$ < T < $4 K), the heat capacity measurement is well described by $C(T)=\gamma T+\beta T^{3}$, were  $\gamma=16$ mJ mole$^{-1}$ K$^{-2}$ \cite{37}. Considering this $\gamma$  value for an evaluation of the Wilson ratio, defined as $R_W=(4\pi^{2}k_B^{2}\chi^P)/[3(g\mu_B)^{2}\gamma]$, the $R_W$ value for Bi$_2$Ir$_2$O$_7$ is 27.1. Since $R_W=1$ for a non-interacting free electron-system, the $R_W$ obtained is indicative of substantial electron-electron correlation in Bi$_2$Ir$_2$O$_7$. As shown in table 1, $\chi_0$ decreases as the Sr content; as the electrical resistivity value as a function of the Sr content remains in the same order, we can roughly assume that $N(E_F)$ has no noticeable variation and $\gamma$ has a small dependence with the Sr content. In this way $R_W$ is expected to decrease as $\chi^P$ and the electron-electron correlation would decrease as the Sr content. We could suppose that such decrease is associated to the presence of Ir$^{5+}$ and that such electronic correlation can be either originated in the SOC (anticipated for all the $5d$ electronic systems) or to short-range correlations of magnetic electrons.

Even though $5d$ orbitals are spatially more extended than $3d$ and $4d$, and the correlations effects are expected to be minimal, there are some iridium-oxide based compounds in which electron correlation seems to play a crucial role in the electronic properties of them. In Sr$_2$IrO$_4$ the SOC is assumed responsible for a Mott ground state of spin driven correlated-electron phenomena \cite{49}; a quantum Hall effect has been anticipated in Na$_2$IrO$_3$ on the basis of spin-orbit interaction and electron correlation \cite{50} and more strongly in the Sr$_{n+1}$Ir$_n$O$_{3n+1}$ series of compounds (n=1, 2 and $\infty$), optical spectroscopy results and first-principles calculations consistently show that large SOC in $5d$ systems could drastically enhance electron correlation effects \cite{51}.

\section{Conclusions}

Magnetic susceptibility measurements on the Bi$_{2-y}$Sr$_{y}$Ir$_2$O$_7$ system indicate a shift towards nonmagnetic Ir$^{5+}$ in samples with $y>0$. The temperature dependence of the magnetic susceptibility is consistent with a trigonal antiprism crystal field splitting ($S=1/2$ for Ir$^{4+}$ and $S=0$ for Ir$^{5+}$). In the 2-300 K temperature range a magnetic transition was  not observed and the small $\theta$ values can be associated with a weak antiferromagnetic coupling in this system. The Bi substitution by Sr in Bi$_{2-y}$Sr$_{y}$Ir$_2$O$_7$, produced a magnetic state whose temperature behavior is not well described by a Curie-Weiss equation. The Bi$_{2-y}$Sr$_{y}$Ir$_2$O$_7$ solid solution is magnetically well described by a modified CW equation in which short-range interaction and magnetic thermal-fluctuations are taken into account. The estimated Wilson-ratio indicates the existence of electron-electron correlations in a paramagnetic state; this correlation seems to extend over the whole range of $y$-composition but decreases with the Sr content.

\section*{Acknowledgements}

We thank to Cecilia Salcedo her help in X-ray diffraction experiments, also technical support from Sra. Irma Vigil is acknowledged. This research has been partially supported by the following UNAM-PAPIIT  Grants:  IN-110210 and IN-118710. RE   thanks  partial support to CONACyT Project 129293 (Ciencia B\'asica), DGAPA-UNAM project IN100711, project BISNANO 2011, and project PICCO 11-7, by El Instituto de Ciencias del  Distrito Federal, Ciudad de M\'exico.

\end{document}